\documentclass[longauth]{aa}  

\usepackage{graphicx}
\usepackage{subfigure}
\usepackage{caption}
\usepackage{txfonts}
\usepackage[breaklinks,colorlinks,citecolor=blue]{hyperref}
\usepackage[flushleft]{threeparttable}
\usepackage{hyperref}
\usepackage{multirow}
\usepackage{lscape}
\usepackage{longtable}

\hypersetup{
  bookmarks=true
  citecolor=blue
  colorlinks   = true, 
  urlcolor     = magenta, 
  linkcolor    = blue, 
  backref = false
  pagebackref = false
}

\begin{document}

   \title{The Gaia-ESO Survey: the present-day radial metallicity distribution of the Galactic disc probed by pre-main-sequence clusters \thanks{Based on observations made with the ESO/VLT, at Paranal Observatory, under program 188.B-3002 (The Gaia-ESO Public Spectroscopic Survey).}}

   \author{L. Spina\inst{\ref{inst1}}
          \and
          S. Randich\inst{\ref{inst2}}
          \and
          L. Magrini\inst{\ref{inst2}}
         \and
         R.~D. Jeffries\inst{\ref{inst9}}
        \and
         E.~D. Friel\inst{\ref{inst24}}
         \and
         G. G. Sacco\inst{\ref{inst2}}
         \and
        E. Pancino\inst{\ref{inst2},\ref{inst3}}
         \and
        R. Bonito\inst{\ref{inst6},\ref{inst7}}
         \and
        L. Bravi\inst{\ref{inst2}}
         \and
         E. Franciosini\inst{\ref{inst2}}
         \and
        A. Klutsch\inst{\ref{inst4}}
        \and
        D. Montes\inst{\ref{inst5}}
         \and
        G. Gilmore\inst{\ref{inst8}}
         \and
         A. Vallenari\inst{\ref{inst10}}
         \and
         T. Bensby\inst{\ref{inst11}}
         \and
         A. Bragaglia\inst{\ref{inst12}}
         \and
         E. Flaccomio\inst{\ref{inst6}}
         \and
         S.~E. Koposov\inst{\ref{inst13},\ref{inst14}}
         \and
         A.~J. Korn\inst{\ref{inst15}}
         \and 
         A.~C. Lanzafame\inst{\ref{inst4},\ref{inst23}}
         \and
         R. Smiljanic\inst{\ref{inst16}}
         \and
         A. Bayo\inst{\ref{inst17}}
         \and
         G. Carraro\inst{\ref{inst18}}
         \and
         A.~R. Casey\inst{\ref{inst8}}
         \and
         M.~T. Costado\inst{\ref{inst19}}
         \and
         F. Damiani\inst{\ref{inst6}}
         \and 
         P. Donati\inst{\ref{inst12}}
         \and 
         A. Frasca\inst{\ref{inst4}}
         \and
         A. Hourihane\inst{\ref{inst8}}
         \and 
         P. Jofr\'e\inst{\ref{inst8},\ref{inst20}}
         \and
         J. Lewis\inst{\ref{inst8}}
         \and
         K. Lind\inst{\ref{inst25}}
         \and
         L. Monaco\inst{\ref{inst21}}
         \and
         L. Morbidelli\inst{\ref{inst2}}
         \and
         L. Prisinzano\inst{\ref{inst6}}
         \and
         S.~G. Sousa\inst{\ref{inst22}}
         \and
         C.~C. Worley\inst{\ref{inst8}}
         \and
         S. Zaggia\inst{\ref{inst10}}
         }

  \institute{Universidade de S\~ao Paulo, IAG, Departamento de Astronomia, Rua do Mat\~ao 1226, S\~ao Paulo, 05509-900 SP, Brasil - \email{lspina@usp.br}\label{inst1}
   \and INAF - Osservatorio Astrofisico di Arcetri, Largo E. Fermi, 5, 50125, Firenze, Italy\label{inst2}
   \and Astrophysics Group, Keele University, Keele, Staffordshire ST5 5BG, United Kingdom\label{inst9}
   \and Department of Astronomy, Indiana University, Bloomington, IN, USA\label{inst24}
   \and ASI Science Data Center, Via del Politecnico SNC, 00133 Roma, Italy\label{inst3}
   \and Dipartimento di Fisica e Astronomia, Sezione Astrofisica, Universit\`{a} di Catania, via S. Sofia 78, 95123, Catania, Italy\label{inst4}
  \and INAF-Osservatorio Astrofisico di Catania, Via S. Sofia 78, I-95123 Catania, Italy\label{inst23}
   \and INAF-Osservatorio Astronomico di Palermo, P.zza del Parlamento 1, I-90134 Palermo, Italy\label{inst6}
   \and Dip. di Fisica e Chimica, Universit\`{a} di Palermo, P.zza del Parlamento 1, I-90134 Palermo, Italy\label{inst7}
   \and Dpto. Astrof\'{i}sica, Facultad de CC. Fisicas, Universidad Complutense de Madrid, E-28040 Madrid, Spain\label{inst5} 
   \and Institute of Astronomy, University of Cambridge, Madingley Road, Cambridge CB3 0HA, United Kingdom\label{inst8}
   \and INAF - Padova Observatory, Vicolo dell'Osservatorio 5, 35122 Padova, Italy\label{inst10}
   \and Lund Observatory, Department of Astronomy and Theoretical Physics, Box 43, SE-221 00 Lund, Sweden\label{inst11}
   \and INAF - Osservatorio Astronomico di Bologna, via Ranzani 1, 40127, Bologna, Italy\label{inst12}
   \and Institute of Astronomy, University of Cambridge, Madingley Road, Cambridge CB3 0HA, United Kingdom\label{inst13}
   \and Moscow MV Lomonosov State University, Sternberg Astronomical Institute, Moscow 119992, Russia\label{inst14}
   \and Department of Physics and Astronomy, Uppsala University, Box 516, SE-751 20 Uppsala, Sweden\label{inst15}
   \and Nicolaus Copernicus Astronomical Center, Polish Academy of Sciences, ul. Bartycka 18, 00-716, Warsaw, Poland\label{inst16}
   \and Instituto de F\'isica y Astronomi\'ia, Universidad de Valparai\'iso, Chile\label{inst17}
   \and European Southern Observatory, Alonso de Cordova 3107 Vitacura, Santiago de Chile, Chile\label{inst18}
   \and Instituto de Astrof\'{i}sica de Andaluc\'{i}a-CSIC, Apdo. 3004, 18080 Granada, Spain\label{inst19}
   \and N\'ucleo de Astronom\'ia, Facultad de Ingenier\'ia, Universidad Diego Portales,  Av. Ejercito 441, Santiago, Chile\label{inst20}
   \and Max-Planck Institut f\"{u}r Astronomie, K\"{o}nigstuhl 17, 69117 Heidelberg, Germany\label{inst25}
   \and Departamento de Ciencias Fisicas, Universidad Andres Bello, Republica 220, Santiago, Chile\label{inst21}
   \and Instituto de Astrof\'isica e Ci\^encias do Espa\c{c}o, Universidade do Porto, CAUP, Rua das Estrelas, 4150-762 Porto, Portugal\label{inst22}
               }

   \date{Received: 17th November 2016; accepted: 9th February 2017}

 
  \abstract
   {The radial metallicity distribution in the Galactic thin disc represents a crucial constraint for modelling disc formation and evolution. Open star clusters allow us to derive both the radial metallicity distribution and its evolution over time.}
   {In this paper we perform the first investigation of the present-day radial metallicity distribution based on [Fe/H] determinations in late type members of pre-main-sequence clusters. Because of their youth, these clusters are therefore essential for tracing the current inter-stellar medium metallicity.}
   {We used the products of the Gaia-ESO Survey analysis of 12 young regions (age$<$100~Myr), covering Galactocentric distances from 6.67 to 8.70 kpc. For the first time, we derived the metal content of star forming regions farther than 500 pc from the Sun. Median metallicities were determined through samples of reliable cluster members. For ten clusters the membership analysis is discussed in the present paper, while for other two clusters (i.e. Chamaeleon~I and Gamma Velorum) we adopted the members identified in our previous works.}
   {All the pre-main-sequence clusters considered in this paper have close-to-solar or slightly sub-solar metallicities. The radial metallicity distribution traced by these clusters is almost flat, with the innermost star forming regions having [Fe/H] values that are 0.10-0.15 dex lower than the majority of the older clusters located at similar Galactocentric radii.}
   {This homogeneous study of the present-day radial metallicity distribution in the Galactic thin disc favours models that predict a flattening of the radial gradient over time. On the other hand, the decrease of the average [Fe/H] at young ages is not easily explained by the models. Our results reveal a complex interplay of several processes (e.g. star formation activity, initial mass function, supernova yields, gas flows) that controlled the recent evolution of the Milky Way.}


\authorrunning{L. Spina et al.}
\titlerunning{The present-day radial metallicity distribution of the Galactic disc probed by pre-main-sequence clusters}

   \maketitle
%

\section{Introduction}
The chemical composition of stars and, in particular, the radial metallicity distribution in the Milky Way disc and their evolution over time, can reveal important clues to the numerous global variables (e.g. star formation activity, initial mass function, supernova yields, radial migration, gas flows) that controlled - and are still governing - the history of baryonic matter in our Galaxy, as has been shown by recent theoretical (e.g. \citealt{Matteucci09,Romano10,Minchev14,Snaith15,Kubryk15a,Kubryk15b,Andrews16,Pezzulli16}) and observational studies (e.g. \citealt{Haywood13,Haywood16,Battistini16,Nissen15,Nissen16,Spina16b,Spina16,Anders16,DOrazi16}). In this context, open star clusters are excellent tracers of the radial metallicity distribution (e.g. \citealt{Friel95,Carraro07,Sestito08,Yong12,Donati15,CantatGaudin16,Netopil16,Reddy16}), allowing us to investigate mechanisms of disc formation and evolution (e.g. \citealt{Magrini09,Magrini15,Jacobson16}). Not only clusters cover large ranges of age and Galactocentric radii, but their distances and ages can be more effectively constrained than in the case of individual field stars. An important exception is represented by the Cepheid variables which are exceptional distance indicators and whose ages are limited within $\sim$20-400 Myr (e.g. \citealt{Bono05}). Nevertheless, the chemical content of open clusters can be more precisely determined than for Cepheids, thanks to the large number of members that can be observed within each association.

In the last few years an increasing number of studies have focused on the metallicity of young open clusters (YOCs) and star forming regions (SFRs) (e.g. \citealt{DOrazi09a,DOrazi09b,DOrazi11,VianaAlmeida09,Biazzo11a,Biazzo11b,Biazzo12a,Biazzo12b,Spina14b,Spina14}). These studies suggest that YOCs, where star formation has ceased, generally share a metallicity close to the solar value; on the other hand, SFRs, in which the molecular gas is still present and the star formation process is still ongoing, seem to be characterised by a slightly lower iron content. In particular, no metal-rich SFRs have been discovered so far. However, these conclusions are mainly based both on small number statistics (typically, 1-5 stars per association), and on [Fe/H] values that are determined from different observations and methods of analysis. In addition, all the stellar associations younger than 100~Myr and with metallicity determinations are located within 500 pc from the Sun. Therefore, the question arises whether the observed metallicities reflect the initial abundance of a giant molecular cloud complex that could have given birth to most of the SFRs and YOCs in the solar vicinity in a common and widespread star formation episode (as proposed by \citealt{Spina14b}) or if these metallicities are the result of a more complex process of chemical evolution that involved a much larger area in the Galactic disc.

New homogeneous studies are needed to i) enable a more homogeneous view of the metal content in nearby YOCs and SFRs; ii) better understand if this metallicity is a peculiarity of the local inter-stellar medium (ISM) or if it is a common pattern of the youngest stars in the entire Galactic disc. In turn, this would allow us to achieve a better understanding of the latest phases of the evolution of the Galactic disc, providing new constraints that theoretical models should take into account.

The Gaia-ESO Survey \citep{Gilmore12,Randich13} will contribute significantly to these studies. It is a large public spectroscopic survey observing all the components of the Galaxy (bulge, thin and thick discs, and halo). The project makes use of the FLAMES spectrograph mounted at the Very Large Telescope to obtain Giraffe and UVES spectra of about 10$^{5}$ stars, including candidate members of 60-70 open clusters. This large body of observations, with homogeneous analysis techniques, will allow us to study the kinematical and chemical abundance distributions in the Milky Way and also to fully sample the age-metallicity-mass/density- Galactocentric distance parameter space within the selected open clusters. 

In this paper, we make use of the advanced products internally released to the Gaia-ESO Survey consortium for five SFRs (Carina Nebula, Chamaeleon~I, NGC2264, NGC6530, and Rho Ophiuchi) and seven YOCs (Gamma Velorum, IC2391, IC2602, IC4665, NGC2547, and NGC2451AB), in order to derive their metal content on a homogeneous scale. The cluster metallicities are derived on the basis of samples of reliable cluster members obtained through the Gaia-ESO dataset. The membership analysis for four SFRs and six YOCs is discussed in this paper, while for Chamaeleon~I and Gamma Velorum we adopted the list of members found in our previous works. It is worth mentioning that we provide for the first time the metallicity determinations of three distant SFRs: the Carina Nebula (distance d$\sim$2.7~kpc), NGC6530 (d$\sim$1.25~kpc), and NGC2264 (d$\sim$0.76~kpc). As we show below, these determinations give new key insights into the present-day metallicity distribution in the thin disc.

In Section~\ref{Data} we describe the Gaia-ESO dataset used for our analysis. Recently, \citet{Spina14,Spina14b} already listed the UVES members of Gamma Velorum and Chamaeleon~I, while the way to identify the likely members belonging to the remaining ten clusters is presented in Section~\ref{membership}. The metal content of each cluster is derived in Section~\ref{Metallicitydeterminations}. In Section~\ref{discussion} we present and discuss the radial metallicity distribution traced by the YOCs and SFRs observed by the Gaia-ESO Survey. Finally, our conclusions are outlined in Section~\ref{conclusions}.

\section{The Gaia-ESO Survey dataset}
\label{Data}
The metallicity determinations of the 12 young stellar associations analysed in this paper are entirely based on the parameters produced by the analysis performed by the Gaia-ESO consortium on the UVES and Giraffe spectra collected by the Gaia-ESO Survey in the period January 2012-July 2014, and on data retrieved from the ESO Archive. The analysis products from both novel and archival spectra have been released internally to the consortium in the iDR4 catalogue stored at the Wide Field Astronomy Unit at Edinburgh University
. In this section, we briefly describe the target selection, the observations, the spectrum analysis, and the available data products of the Gaia-ESO Survey.
\subsection{Target selection and observations}
The Gaia-ESO Survey observations were performed at the Very Large Telescope using the multi-object optical spectrograph FLAMES \citep{Pasquini02} in Medusa feeding mode, allowing the simultaneous allocation of UVES high-resolution (R$\sim$47~000) and Giraffe intermediate-resolution (R$\sim$17~000) fibres. This system permits the simultaneous allocation of eight fibres feeding the UVES spectrograph plus 132 additional fibres feeding Giraffe.

The UVES observations of SFRs and YOCs were performed with the 580nm setup for F-,G-, and K-type stars, while the 520nm setup was employed for the observations of earlier-type targets (not considered here). The Giraffe spectra of late-type young stars (from F to M-type) were acquired with the HR15N setup. For warmer stars additional setups have been used, but these sources are not considered in this paper. We note that the 580nm and HR15N setups contain the lithium line at 6708~$\AA$ that is an important diagnostic of youth for late-type stars.

%
\subsection{Data reduction and analysis}
Both the Gaia-ESO Survey and the archival spectra, have been reduced and analysed by the Gaia-ESO consortium in a homogeneous way.
UVES data reduction is carried out using the FLAMES-UVES ESO public pipeline \citep{Modigliani04}. This procedure and the determination of the radial velocities (RVs) from UVES spectra are described in \citet{Sacco14}. A dedicated pipeline has been developed by the Cambridge Astronomical Survey Unit (CASU) to reduce the Giraffe spectra and to derive RVs and rotational velocity values (v$sin$i). Details on this approach are reported in Gilmore et al. (2016, in prep.), but a summary of the Giraffe data reduction is given by \citet{Jeffries14}.

Different Working Groups (WGs) contributed to the spectrum
analysis through the accomplishment of distinct duties. Groups WG10, WG11, and WG12 perform the analysis of the spectra of cool (FGKM-type) targets observed in the clusters discussed in this paper. WG10 is dedicated to the analysis of Giraffe spectra of stars in the Milky Way field and in intermediate-age and old cluster fields, WG11 analyses the corrispective UVES spectra observed in the same regions, while 
WG12 focuses on  the analysis of SFRs and YOCs. The detailed analysis procedure performed by these WGs is described in \citet{Smiljanic14} and \citet{Lanzafame15}. Parameter and abundance homogenisation across WGs is then performed by WG15. The calibration and homogenisation strategies are described in \citet{Pancino16} and \citet{Hourihane17}. A brief description of the analysis carried out on all the spectra observed in the fields of YOCs and SFRs is also reported in \citet{Spina14}.

The recommended parameters produced by this analysis, reported in the GESiDR4Final catalogue and used in this work, include the stellar parameters: effective temperatures (T$_{\rm eff}$), surface gravities (log~$g$), metallicities ([Fe/H]), microturbulent velocities ($\xi$). The [Fe/H] values have been derived together with the stellar parameters and are based on the Fe I lines. The GESiDR4Final catalogue also includes the equivalent widths (or upper limits) of the lithium line at 6707.8~$\AA$ (EW(Li)). In iDR4, EW(Li) measurements are provided by WG12. These values are corrected for the contamination from blending of adjacent lines (see \citealt{Lanzafame15}). The Gaia-ESO Survey also determines the $\gamma$ index \citep{Damiani14}, which acts as a gravity indicator for late-type stars \citep{Damiani14,Prisinzano16}.


\section{Membership analysis}
\label{membership}

The Gaia-ESO Survey provides a variety of commonly used spectroscopic tracers (e.g. lithium abundances, radial velocities, surface gravities) that can be used to assess membership to pre-main-sequence clusters, such as SFRs and YOCs (e.g. \citealt{Spina14,Spina14b}). 
For instance, it is well known that the strength of the lithium line at 6708~$\AA$ is an extremely reliable indicator of membership for young G and K-type stars, since they have not had time to significantly deplete the element in their atmospheres (e.g. \citealt{ZapateroOsorio02,Jeffries03}). In fact, all the members of SFRs are expected to have retained their primordial reservoirs of Li, while in YOCs only the M-type stars have partially or completely burned this element. In addition to the log~$g$ values, that are very useful to identify background giant contaminants, the Gaia-ESO Survey also derives the $\gamma$ index for all Giraffe targets observed with HR15N. This index, defined from strongly gravity-sensitive lines by \citet{Damiani14}, is an efficient gravity indicator, allowing a clear separation between the low gravity giants ($\gamma$$\gtrsim$1) and the higher gravity main-sequence and pre-main-sequence stars ($\gamma$$\lesssim$1) for spectral types later than G (see also \citealt{Prisinzano16}). The empirical calibration of the $\gamma$ index done by \citet{Damiani14} on the Gamma Velorum cluster members showed that a threshold value $\gamma$=1.02 quite clearly separates the M- and K-type pre-main-sequence stars from the field giant contaminants. 

In this section we describe the criteria and the procedures adopted to compile a list of reliable members for ten SFRs and YOCs. The membership of Chamaeleon~I and Gamma Velorum has been already analysed by \citet{Spina14,Spina14b}. As detailed below, we adopted slightly different criteria of membership depending on the class of the cluster: SFR or YOC. We note that we generally did not use radial velocities: criteria of membership based on the kinematics, even if commonly employed for older clusters where lithium cannot be used, ensure neither the removal of all the contaminants nor the inclusion of all the members, given the complex kinematic structures that these young regions may have (e.g. \citealt{Jeffries14,Sacco15,Cottaar15,Sacco17}). 

\begin{figure*}
\centering
\includegraphics[width=18cm]{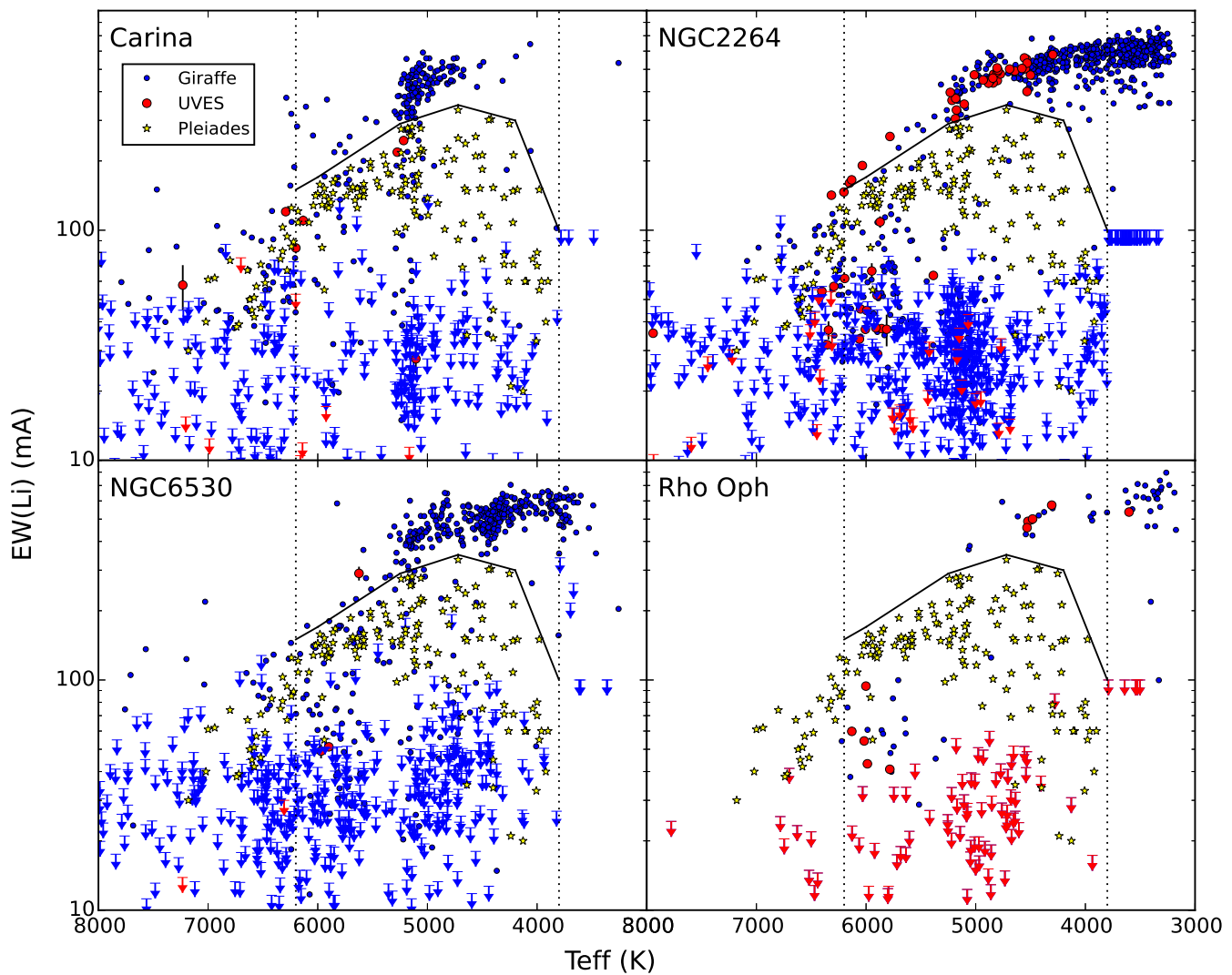} \caption{Lithium equivalent widths (EWs) as a function of the T$_{\rm eff}$ for the candidate members of Carina Nebula, NGC2264, NGC6530, and Rho Ophiuchi. The red symbols identify the UVES targets. Most of the
Li detections in UVES spectra have uncertainties associated to their EWs smaller than the data points. The Giraffe targets are represented by the blue symbols. The solid line, that in the range between 3800 and 6200~K corresponds to the upper-envelope of the Pleiades distribution (yellow stars), denotes the boundary between the locus of the cluster members (above the line) and contaminants (below the line). The dotted lines delimit the temperature range considered for the membership analysis based on Li.}
\label{lithiumSFR}
\end{figure*}

We emphasise that, since the main goal of the present paper is to determine a robust value of the cluster metallicities, our membership analysis is conservative; in other words, we aim to identify the stars that have the highest probability of being members of the clusters and not to provide a complete list of candidate members. 
The secure members singled out through our analysis are listed in Table~\ref{full_parameters}, available in electronic form, together with their fundamental parameters and the other quantities (e.g. EW(Li) and $\gamma$ index) used for their identification.

\subsection{Star forming regions (age~$\lesssim$~5~Myr)}
In this subsection we considered an initial sample of all the stars observed in the SFR fields and with an available T$_{\rm eff}$ determination in the iDR4 catalogue.
This catalogue also lists log~$g$ values for a fraction of these stars. We used these determinations to identify and reject all the secure giant stars. Namely, we considered as giant contaminants all the stars having log~$g$ values that are lower than 3.5~dex, taking into account also the error bars (i.e. log~$g$~+~$\sigma_{log~g}$~$<$~3.5~dex). In addition, we used the $\gamma$ index to identify additional giant stars in these fields: all the stars with T$_{\rm eff}$$<$5200~K and $\gamma$$>$1.02 are considered as giant contaminants. As for the rejection based on the log~$g$, we took into account the uncertainties in both T$_{\rm eff}$ and $\gamma$.

\begin{figure*}
\centering
\includegraphics[width=18cm]{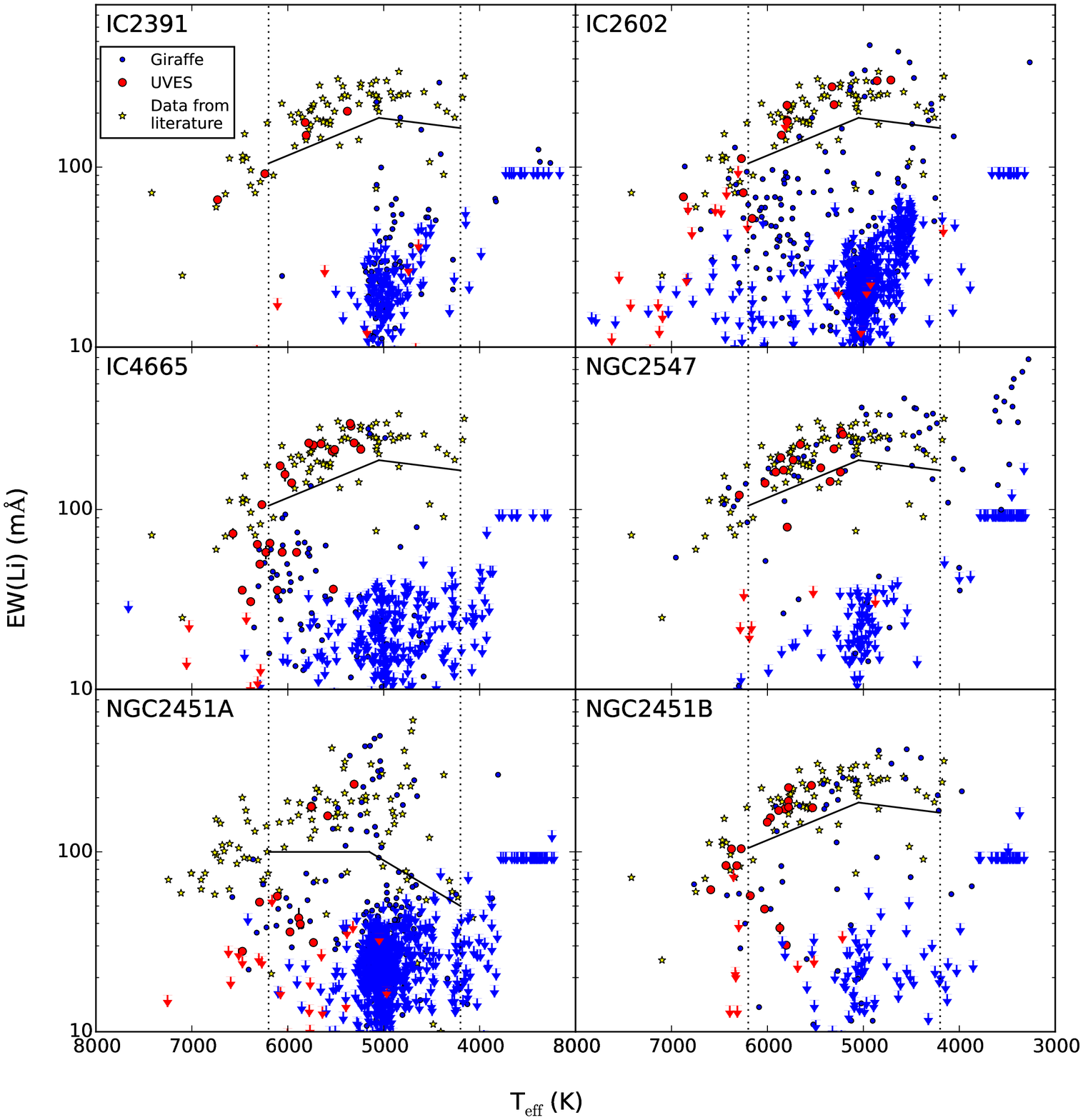} \caption{As Fig.~\ref{lithiumSFR}, but for the YOC candidate members. The clusters with ages $\leq$50~Myr are compared with literature data from IC2391 \citep{Randich01}, IC2602 \citep{Randich97,Randich01}, and IC4665 \citep{Martin97,Jeffries09b}, while the NGC2451A cluster has been compared with data from Alpha Persei \citep{Balachandran11} and from NGC2451A itself \citep{Hunsch04b}.}
\label{lithiumYOC}
\end{figure*}

The equivalent width of the lithium line at 6708~$\AA$ is an excellent youth indicator for G and K-type stars and it can be reliably used to identify the members of SFRs from a sample contaminated by field stars.
The graphs in Fig.~\ref{lithiumSFR} show the EW(Li) of the remaining candidate members plotted as a function of their stellar T$_{\rm eff}$. Together with the Gaia-ESO data, we overplotted the measurements of the Pleiades (age$\sim$125-130~Myr, \citealt{Stauffer98}) obtained from the literature \citep{Soderblom93,Jones96b}. Typically, Pleiades members have a slightly lower amount of Li in their atmospheres relative to stars younger than 5~Myr. Thus, we used the upper envelope of the Pleiades distribution (black line) to assess the membership of the stars in the SFR samples with T$_{\rm eff}$ ranging from 3800 to 6200~K: all the stars in this T$_{\rm eff}$ range and lying above the Pleiades distribution are very likely younger than $\sim$100~Myr, therefore they have been considered as cluster members. Outside of this range of T$_{\rm eff}$, the membership based on Li is more uncertain, since lithium depletion has a different dependence on age (e.g. brown dwarfs below 0.065~M$_{\sun}$ do not burn lithium, while hotter stars take much longer timescales for a significant depletion of the element). For this reason, we discarded all the stars cooler than 3800~K and hotter than 6200~K. Therefore, all the stars lying above the black lines in the plots of Fig.~\ref{lithiumSFR} have been flagged as highly probable members of the SFRs. 

\subsection{Young open clusters (30~$\lesssim$~age~$\lesssim$~80~Myr)}
Here we discuss the membership analysis of the young open clusters IC2391, IC2602, IC4665, NGC2547, and NGC2451 that have ages ranging approximately from 30 to 80~Myr. It is well known that NGC2451 is composed of two clusters located at different distances along the same line of sight: NGC2451A and NGC2451B \citep{Maitzen86,Roser94,Platais96}. The two populations are distinguishable by different radial velocity distributions: based on Gaia-ESO data, \citet{Franciosini17} found that the distribution of NGC2451A is centred at RV$_{A}$=23.41$\pm$0.12~km~s$^{-1}$ with a dispersion $\sigma_{\rm A}$=0.45$\pm$0.28~km~s$^{-1}$, while that of NGC2451B is at RV$_{B}$=15.22$\pm$0.06~km~s$^{-1}$ with a dispersion $\sigma_{B}$=0.15$\pm$0.10~km~s$^{-1}$. We employed the RV values used by Franciosini et al. and their results to separate the stars observed in the NGC2451AB fields into two samples associated to the two clusters. Namely, all the stars with RV~$>$~17.27~km~s$^{-1}$ are considered as NGC2451A candidate members, while the others could belong to NGC2451B. 

Similarly to the SFRs, we initially considered for this analysis all the stars whose UVES and Giraffe spectra have been analysed by the Gaia-ESO consortium and that have T$_{\rm eff}$ determinations. As a first step, we rejected all the possible giant field stars using the same methods adopted for the SFRs based on the log~$g$ values and the $\gamma$ index.

In Fig.~\ref{lithiumYOC} we show the EW(Li) at 6708~$\AA$ as a function of T$_{\rm eff}$ for all the objects that have not been rejected as contaminants.
As for the SFRs, in Fig.~\ref{lithiumYOC} we compared the Gaia-ESO determinations with datasets taken from the literature. Instead of overplotting the measurements of the Pleiades, the YOCs have been compared with clusters of similar ages. The YOCs IC2391, IC2602, IC4665, NGC2547, and NGC2451B have ages younger than 50~Myr and have been compared with the combined data from IC2391 \citep{Randich01}, IC2602 \citep{Randich97,Randich01}, and IC4665 \citep{Martin97,Jeffries09b}. We used the literature values from all of these studies together to define a threshold, below which stars will be considered non-members. Namely, the black solid lines in Fig.~\ref{lithiumYOC} trace the lower envelope of the bulk of cluster members taken from the literature: all the stars of our sample lying above the solid lines have been considered as members. Similarly, NGC2451A that has a slightly older age (about 50-80~Myr) is compared with data from the 60 Myr old cluster Alpha Persei \citep{Balachandran11} and from NGC2451A itself \citep{Hunsch04b}. 

Since all the clusters in Fig.~\ref{lithiumYOC} are older than 30~Myr, their cooler members might have significantly depleted the lithium element. This implies that the strength of the lithium line is not a secure indicator of membership for late-type stars (i.e. $\lesssim$4200~K; e.g., \citealt{Barrado04,Jeffries05,Manzi08}). In addition, as discussed above, the membership based on Li is unreliable for stars with spectral types hotter than F. For this reason, our membership analysis for the YOCs is limited only to stars with T$_{\rm eff}$ ranging from 4200 to 6200~K in order to avoid possible contaminations from field stars.

All the members of the clusters discussed in this section and the parameters used for the membership analysis are listed in Table~\ref{full_parameters}.
A subsample of the secure members has been considered for the metallicity determination of each cluster, as detailed in the next section. 

\begin{table*}
\begin{center}
\tiny
\begin{threeparttable}
\caption{\label{full_parameters} Stellar parameters, EWs of the lithium line at 6707.8~$\AA$ and $\gamma$ indexes of the cluster members. Values from the Gaia-ESO Survey iDR4 catalogue.}
\begin{tabular}{ccccccccccc} 
\hline\hline 
CNAME & RA & DEC & Cluster & Grating & T$_{\rm eff}$ &  log~$g$ & [Fe/H] &  $\xi$ & EW(Li) & $\gamma$ index \\
 & (J2000) & (J2000) & &  & [K] & [dex] & [dex] & [km~s$^{-1}$] & [m$\AA$] &  \\ \hline
06392506+0942515 & 06 39 25.06 & +09 42 51.5 & NGC2264 & HR15N & 4704 $\pm$ 267 & --- & --- & --- & 479.8$\pm$11.4 & 0.916$\pm$0.011 \\
06392550+0931394 & 06 39 25.50 & +09 31 39.4 & NGC2264 & HR15N & 4272 $\pm$ 140 & --- & --- & --- & 487.7$\pm$7.9 & 0.881$\pm$0.005 \\
... & ... & ... & ... & ... & ... & ... & ... & ... & ... & ... \\ \hline
\end{tabular}
\begin{tablenotes}
\item Note: The full version is available at the CDS. 
\end{tablenotes}
\end{threeparttable}
\end{center}
\end{table*}

%
\section{Metallicity determinations}
\label{Metallicitydeterminations}
The metallicity determinations for the clusters analysed in this paper are performed considering a fraction of the UVES and Giraffe secure members identified as detailed in Section~\ref{membership}. Specifically, we only considered stars in the 4200-6200~K temperature range, the [Fe/H] values of which should be highly reliable. This further selection is necessary mainly because of the intrinsic difficulties in the spectral analysis of cooler stars and because of the uncertainties in the membership of the hotter targets. By doing so we ensure a homogeneity in the metallicity determination of all the clusters, also because the choice of this range can be the same for both SFRs and YOCs. 

\begin{figure}
\centering
\includegraphics[width=9cm]{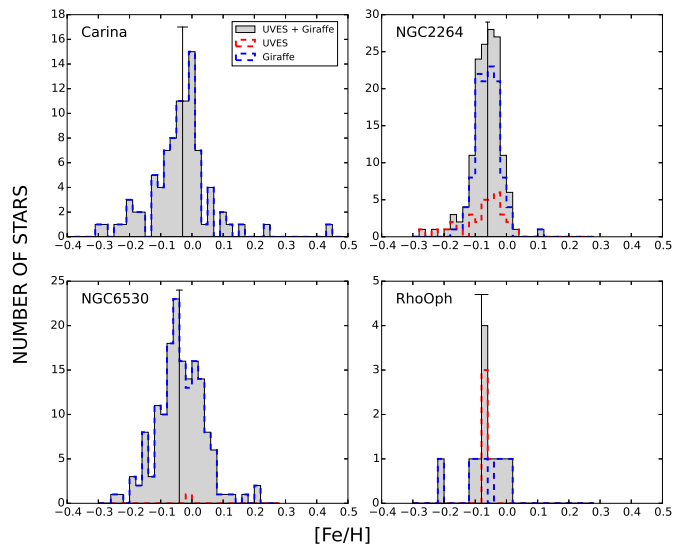} \caption{Metallicity distributions for the SFR members. The filled grey histograms include the [Fe/H] values based on both the UVES and Giraffe spectra, while the red and blue dashed lines show respectively the distributions of the UVES and Giraffe targets alone. Median values and the median absolute deviations of the UVES+Giraffe samples are also represented in the graphs.}
\label{metSFR}
\end{figure}

\begin{figure}
\centering
\includegraphics[width=9cm]{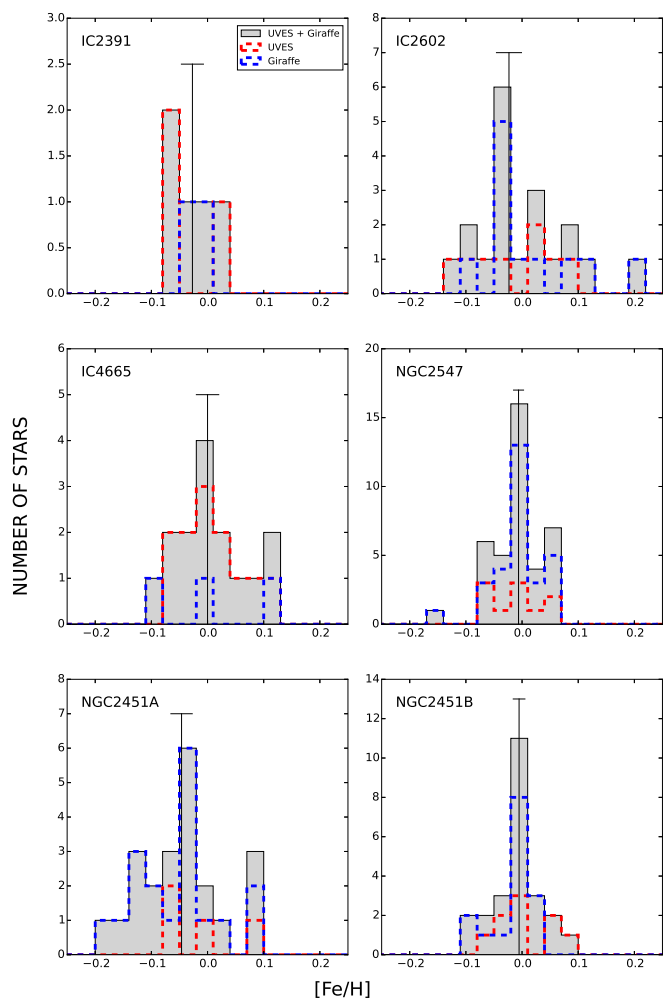} \caption{As Fig.~\ref{metSFR}, but for the YOC members.
}
\label{metYOC}
\end{figure}

The metallicity distributions of the SFRs and YOCs are plotted in Figs.~\ref{metSFR} and \ref{metYOC}, respectively. The metallicity determinations of pre-main-sequence stars could be affected by several issues, such as high stellar rotation or strong emission lines, that can complicate the spectral analysis. Due to the young age of these clusters, it is also possible that some of these stars have been slightly enriched in metals by a recent episode of planetary engulfment (see \citealt{Laughlin97,Spina15}). Furthermore, a fraction of the members observed by the Survey may be components of binary systems and their spectra could be contaminated by a secondary star. All these aspects may explain the presence in Figs. \ref{metSFR} and \ref{metYOC} of some stars with [Fe/H] values that are discrepant ($>$0.10~dex) relative to the peak metallicity distributions. For this reason, in order to reduce the impact of these outliers on our analysis, hereafter we adopt as cluster metallicities the median of the [Fe/H] values of the total distribution that includes both the UVES and Giraffe targets (i.e. $\widetilde{\rm [Fe/H]}_{U+G}$ in the last column in Table~\ref{metallicity}). Similarly, the uncertainty associated to the $\widetilde{\rm [Fe/H]}$ values is the standard error of the median. The results are listed in Table~\ref{metallicity} together with other parameters (i.e. age, distance from the Sun, Galactocentric radius) that characterise each cluster. We noted that the median metallicity values obtained through the UVES and Giraffe targets separately are consistent within the uncertainties, as shown in Table~\ref{metallicity}. In the last columns we reported in brackets the final number of stars for each cluster and each spectrograph that we used for the metallicity determinations. 

The stars observed in the Carina and NGC2264 fields are part of different sub-clusters: Trumpler~14 and Trumpler~16 in Carina \citep{Hur12} and S MON and CONE in NGC2264 \citep{Sung09}. Based on the spatial distribution of these targets, we explored the possibility that these clusters within the same regions have dissimilar metallicities. However, we did not find any significant difference or spatial heterogeneities in their metal contents.

\begin{table*}
\begin{center}
\begin{tiny}
\begin{threeparttable}
\caption{\label{metallicity} Metallicities of the SFRs and YOCs whose membership is discussed in this paper. The associations are sorted by Galactocentric radii.}
\begin{tabular}{c|cc|cccc|ccc} 
\hline\hline 
Clusters & l & b & Age & Distance & Ref. & $R_{gal}^{*}$ & $\widetilde{\rm [Fe/H]}_{G}^{**}$ & $\widetilde{\rm [Fe/H]}_{U}^{**}$ & $\widetilde{\rm[Fe/H]}_{U+G}$\\
& (J2000) & (J2000) & [Myr] & [kpc] &  & [kpc] & [dex] & [dex] & [dex] \\ \hline
\multicolumn{10}{c}{Star forming regions} \\\hline
NGC6530 & 6.082 & $-$1.331 & 1-2 & 1.5$\pm$0.3 & 1 & 6.5$\pm$0.3 & $-$0.041$\pm$0.009 (161) & 0.00 (1) & $-$0.041$\pm$0.009 \\
Carina & 287.408 & $-$0.577 & 1-3 & 2.7$\pm$0.3 & 2,3,4,5 & 7.64$\pm$0.02 & $-$0.030$\pm$0.016 (91) & --- (0) & $-$0.030$\pm$0.016 \\
Rho Oph & 353.686 & +17.687 & 2-3 & 0.13$\pm$0.01 & 6,7 & 7.88$\pm$0.01 & $-$0.07$\pm$0.03 (7) & $-$0.080$\pm$0.005 (4) & $-$0.08$\pm$0.02 \\
NGC2264 & 202.935 & +2.195 & 1-3 & 0.76$\pm$0.09 & 8,9 & 8.70$\pm$0.08 & $-$0.059$\pm$0.004 (114) & $-$0.07$\pm$0.02 (34) & $-$0.060$\pm$0.006\\ \hline
\multicolumn{10}{c}{Young open clusters} \\\hline
IC4665 & 30.619 & +17.081 & 30$\pm$5 & 0.36$\pm$0.01 & 10,11 & 7.71$\pm$0.01 & 0.00$\pm$0.06 (3) & 0.00$\pm$0.02 (12) & 0.00$\pm$0.02\\
IC2602 & 289.601 & $-$4.906 & 30$\pm$5 & 0.15$\pm$0.01 & 12,13 & 7.95$\pm$0.01 & $-$0.02$\pm$0.03 (11) & 0.00$\pm$0.03 (8) & $-$0.02$\pm$0.02\\
IC2391 & 270.362 & $-$6.839 & 55$\pm$5 & 0.16$\pm$0.01 & 14,15 & 8.00$\pm$0.01 & $-$0.013$\pm$0.012 (2) & $-$0.05$\pm$0.03 (3) & $-$0.03$\pm$0.02 \\
NGC2547 & 264.435 & $-$8.625 & 35$\pm$5 & 0.36$\pm$0.02 & 16,17,18 & 8.04$\pm$0.01 & $-$0.001$\pm$0.010 (29) & $-$0.015$\pm$0.019 (10) & $-$0.006$\pm$0.009\\
NGC2451A & 252.575 & $-$7.298 & 50-80 & 0.21$\pm$0.01 & 19 & 8.06$\pm$0.01 & $-$0.05$\pm$0.02 (18) & $-$0.04$\pm$0.04 (4) & $-$0.046$\pm$0.019\\
NGC2451B & 252.050 & $-$6.726 & 50$\pm$10 & 0.37$\pm$0.01 & 19 & 8.12$\pm$0.01 & $-$0.006$\pm$0.012 (15) & 0.002$\pm$0.017 (9) & $-$0.005$\pm$0.010\\ \hline
\end{tabular}
\begin{tablenotes}
\item References for cluster ages and distances: 1) \citet{Prisinzano05}; 2) \citet{Smith00}; 3) \citet{DeGioia-Eastwood01}; 4) \citet{Smith06}; 5) \citet{Hur12}; 6) \citet{Mamajek08}; 7) \citet{Erickson11}; 8) \citet{Sung97} ; 9) \citet{Sung04}; 10) \citet{Manzi08}; 11) \citet{Cargile10}; 12) \citet{vanLeeuwen09}; 13) \citet{Stauffer97}; 14) \citet{Barrado99}; 15) \citet{Barrado04};16) \citet{Jeffries05}; 17) \citet{Lyra06}; 18) \citet{Naylor06}; 19) \citet{Hunsch03}.

\item *) We adopted $R_{gal}$ = 8.0 kpc for the Sun.
\item **) We reported in brackets the final number of stars for each cluster and each spectrograph that we used for the metallicity determinations.
\end{tablenotes}
\end{threeparttable}
\end{tiny}
\end{center}
\vspace{-0.4cm}
\end{table*}

\begin{table}
\begin{tiny}
\caption{\label{newFe} Stellar members of Chamaeleon~I and Gamma Velorum used for the metallicity determinations of the two clusters by Spina et al (2014ab). The listed atmospheric  parameters are those released in iDR4.}
\begin{tabular}{c|cccc} 
\hline\hline 
2MASS & T$_{\rm eff}$ & log~$g$ & [Fe/H] & $\xi$ \\
NAME & [K] & [dex] & [dex] & [km s$^{-1}$] \\ \hline
\multicolumn{5}{c}{Chamaeleon~I} \\\hline
11022491-7733357 & 4543$\pm$174 & 4.51$\pm$0.16 & $-$0.07$\pm$0.13 & --- \\
11045100-7625240 & 4575$\pm$111 & 4.47$\pm$0.15 & $-$0.08$\pm$0.14 & 1.79$\pm$0.15 \\
11064510-7727023 & 4316$\pm$104 & 4.62$\pm$0.13 & 0.00$\pm$0.12 & --- \\
11114632-7620092 & 4630$\pm$162 & 4.5$\pm$0.18 & $-$0.06$\pm$0.14 & --- \\
11124299-7637049 & 4567$\pm$152 & 4.31$\pm$0.17 & $-$0.11$\pm$0.13 & --- \\
11291261-7546263 & 4823$\pm$98 & 4.5$\pm$0.15 & $-$0.07$\pm$0.15 & --- \\ \hline
\multicolumn{5}{c}{Gamma Velorum} \\\hline
08110285-4724405 & 5233$\pm$81 & 4.47$\pm$0.15 & $-$0.03$\pm$0.12 & --- \\
08095967-4726048 & 5214$\pm$117 & 4.34$\pm$0.27 & 0.00$\pm$0.13 & --- \\
08095427-4721419 & 5884$\pm$93 & 4.44$\pm$0.15 & 0.10$\pm$0.12 & 1.61$\pm$0.15 \\
08094221-4719527 & 5128$\pm$142 & 4.49$\pm$0.13 & $-$0.03$\pm$0.13 & --- \\
08093304-4737066 & 5640$\pm$127 & 4.26$\pm$0.15 & $-$0.01$\pm$0.13 & 1.82$\pm$0.15 \\
08092627-4731001 & 5220$\pm$93 & 4.52$\pm$0.12 & $-$0.05$\pm$0.11 & --- \\
08090850-4701407 & 6650$\pm$201 & 4.10$\pm$0.13 & $-$0.06$\pm$0.14 & --- \\
08091875-4708534 & 6708$\pm$132 & 3.94$\pm$0.16 & $-$0.04$\pm$0.13 & --- \\ \hline
\end{tabular}
\end{tiny}
\vspace{-0.4cm}
\end{table}

\begin{table*}
\begin{center}
\begin{threeparttable}
\caption{\label{metallicity_literature} Metallicities of the associations observed by the Gaia-ESO Survey whose membership is known from the literature. The associations are sorted by Galactocentric radii.}
\begin{tabular}{c|cc|cccc|c} 
\hline\hline 
Clusters & l & b & Age & Distance & Ref. & $R_{gal}^{*}$ & $\widetilde{\rm [Fe/H]}^{**}$\\
& (J2000) & (J2000) & [Myr] & [kpc] &  & [kpc] & [dex] \\ \hline
\multicolumn{8}{c}{Star forming region} \\\hline
Chamaeleon I & 297.1559 & $-$15.617 & 2-5 & 0.16$\pm$0.01 & 1,2 & 7.93$\pm$0.01 & $-$0.070$\pm$0.017 (6) \\ \hline
\multicolumn{8}{c}{Young open cluster} \\\hline
Gamma Velorum & 262.8025 & $-$07.686 & 10-20 & 0.35$\pm$0.01 & 3 & 8.05$\pm$0.01 & $-$0.03$\pm$0.02 (8) \\ \hline
\multicolumn{8}{c}{Intermediate-age open clusters} \\\hline
Berkeley 81 & 34.51 & $-$2.07 & 860$\pm$100 & 3.50$\pm$0.12 & 5,6 & 5.49$\pm$0.07 & 0.22$\pm$0.02 (13) \\
NGC 6005 & 325.8 & $-$3.00 & 1200$\pm$300 & 2.7$\pm$0.5 & 10 & 6.0$\pm$0.3 & 0.155$\pm$0.007 (12) \\
Trumpler 23 & 328.8 & $-$0.50 & 800$\pm$100 & 2.2$\pm$0.2 & 4 & 6.25$\pm$0.15 & 0.140$\pm$0.012 (10) \\
NGC 6705 & 27.31 & $-$2.78 & 300$\pm$50 & 2.0$\pm$0.2 & 12 & 6.34$\pm$0.16 & 0.080$\pm$0.013 (27) \\
Pismis 18 & 308.2 & 0.30 & 1200$\pm$400 & 2.2$\pm$0.4 & 10 & 6.85$\pm$0.17 & 0.105$\pm$0.011 (6) \\
Trumpler 20 & 301.48 & 2.22 & 1500$\pm$150 & 3.5$\pm$0.3 & 13 & 6.86$\pm$0.02 & 0.120$\pm$0.010 (42) \\
Berkeley 44 & 53.2 & +3.33 & 2900$\pm$300 & 2.2$\pm$0.4 & 4 & 6.91$\pm$0.12 & 0.18$\pm$0.03 (4) \\
NGC 4815 & 303.63 & $-$2.10 & 570$\pm$70 & 2.50$\pm$0.15 & 9 & 6.94$\pm$0.04 & $-$0.03$\pm$0.03 (5) \\
NGC 6802 & 55.3 & 0.92 & 1000$\pm$100 & 2.3$\pm$0.2 & 4 & 6.96$\pm$0.07 & 0.100$\pm$0.007 (8) \\
NGC 6633 & 36.0 & 8.3 & 630$\pm$100 & 0.36$\pm$0.02 & 11 & 7.71$\pm$0.01 & $-$0.06$\pm$0.02 (11) \\
NGC 3532 & 289.6 & 1.35 & 300$\pm$100 & 0.49$\pm$0.01 & 8 & 7.85$\pm$0.01 & $-$0.025$\pm$0.013 (2) \\
NGC 2516 & 273.8 & $-$15.8 & 163$\pm$40 & 0.36$\pm$0.02 & 7 & 7.98$\pm$0.01 & $-$0.080$\pm$0.016 (15) \\ \hline

\end{tabular}
\begin{tablenotes}
\item References for cluster ages and distances: 1) \citet{Whittet97}; 2) \citet{Luhman07}; 3) \citet{Jeffries09}; 4) \citet{Jacobson16}; 5) \citet{Donati14}; 6) \citet{Magrini15}; 7) \citet{Sung02}; 8) \citet{Clem11}; 9) \citet{Friel14}; 10) \citet{Piatti97}; 11) \citet{Jeffries02}; 12) \citet{CantatGaudin14}; 13) \citet{Donati14b}.
\item *) We adopted $R_{gal}$ = 8.0 kpc for the Sun.
\item **) The metallicity values are the median of the [Fe/H] reported in iDR4 for the stellar members of the clusters. We reported in brackets the number of stars that we used for the metallicity determinations.

\end{tablenotes}
\end{threeparttable}
\end{center}
\vspace{-0.4cm}
\end{table*}

The first set of data internally released by the Gaia-ESO Survey included parameters for the stars observed in Chamaeleon~I (age$\sim$3~Myr) and Gamma Velorum (age$\sim$10-20~Myr). This dataset has been employed by \citet{Spina14,Spina14b} to identify the members of these associations and to determine the mean cluster metallicities. 
We derived the median metallicities of the two regions using iron abundances recommended in iDR4 for the members with T$_{\rm eff}$$>$4200~K already identified in our previous papers.
The atmospheric parameters of these stars are reported in Table~\ref{newFe}. The new median metallicity found for Chamaeleon~I is $-$0.07$\pm$0.04~dex, while for Gamma Velorum it is $-$0.03$\pm$0.02~dex. 
These values are consistent with those provided by \citet{Spina14,Spina14b}: $-$0.08$\pm$0.04 and $-$0.057$\pm$0.018~dec for Chamaeleon I and Gamma Velorum, respectively. The median metallicities of these two clusters are reported in Table~\ref{metallicity_literature} together with the cluster ages and Galactocentric radii.

In the next Section we will compare the metal content of SFRs and YOCs with the metallicities of all the older clusters (0.100~$\lesssim$~age~$\lesssim$~3.0~Gyr) observed by the Gaia-ESO Survey and analysed in iDR4. All the iDR4 atmospheric parameters of the members of these intermediate age clusters have been published by \citet{Jacobson16}. We used the [Fe/H] values listed in that paper to calculate their median metallicities . These values are also reported in Table~\ref{metallicity_literature}.

\section{Discussion}
\label{discussion}
\begin{figure*}
\includegraphics[width=18cm]{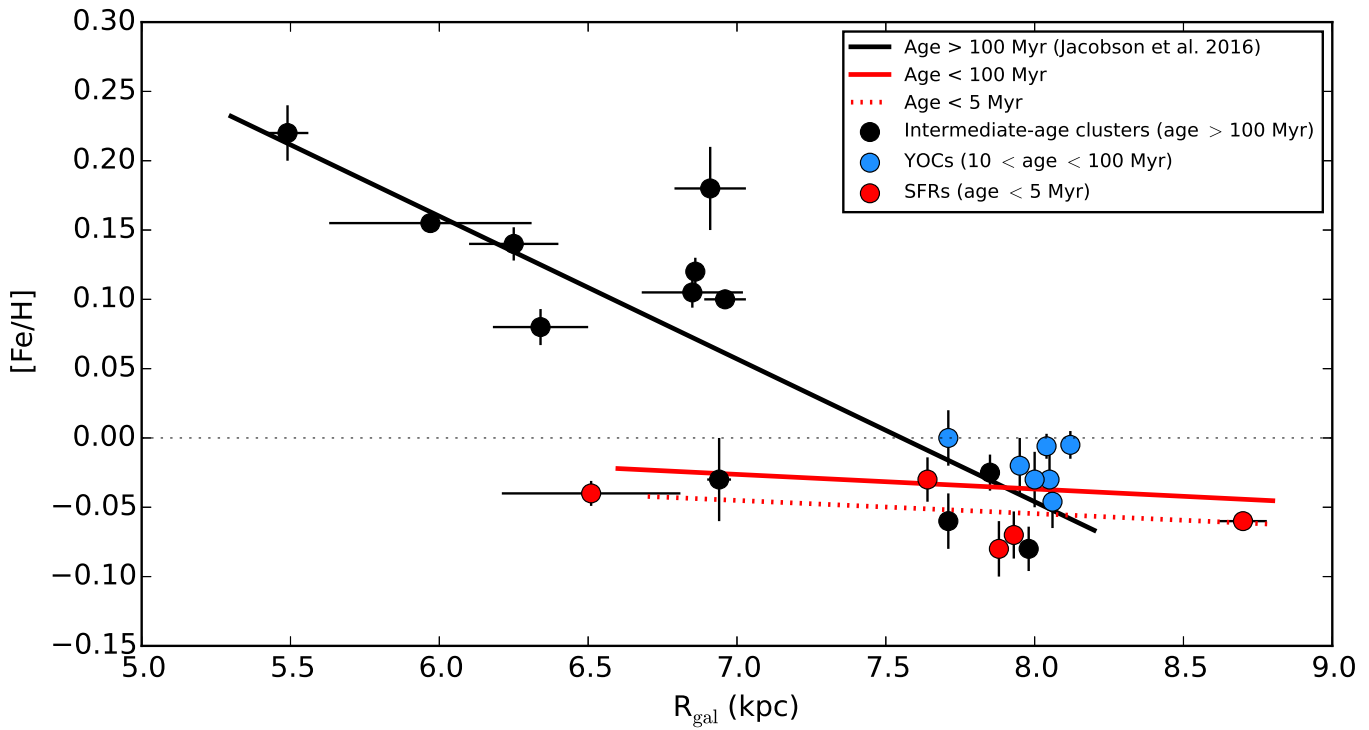} \caption{Radial metallicity distribution of all the clusters included in the iDR4 catalogue. Their Galactocentric radii and metallicities are reported in Tables~\ref{metallicity} and \ref{metallicity_literature}. Red circles indicate star forming regions, blue circles denote young clusters in the age range between 10 and 100~Myr, while intermediate age clusters are shown as black symbols.}
\label{gradients_obs}
\end{figure*}

In Fig.~\ref{gradients_obs} we show the radial distribution of our sample targets along with Chamaeleon~{\sc i} and the Gamma Vel clusters, as well as the older clusters from the Gaia-ESO Survey iDR4 presented by \citet{Jacobson16}. All the clusters shown in the plot are on the same metallicity scale, since homogenisation of metallicities has been performed by Gaia-ESO \citep{Pancino16,Hourihane17}. The figure shows two main features. First, although some dispersion is present at the solar radius, the five star forming regions and the seven young open clusters (age $\leq 100$~Myr) have very similar average metallicities [Fe/H]$_{\rm SFR}=-0.056\pm0.018$ and [Fe/H]$_{\rm YOC}=-0.020\pm0.016$. In particular, none of these associations appear to be metal rich. This result confirms and extends, both to a larger number of clusters (and cluster members) and to larger distances from the Sun, the early findings by \citet{Biazzo11a} and \citet{Spina14b}. Both studies noted that, while open clusters in the solar vicinity ($\lesssim$500 pc) cover a large range in metallicity values (i.e. $-$0.2$\lesssim$[Fe/H]$\lesssim$0.3~dex), the youngest stars ($\lesssim$100~Myr) within that volume are restricted to the lowest metallicities (i.e. $-$0.2$\lesssim$[Fe/H]$\lesssim$0.0~dex). In particular,  \citet{Spina14b} hypothesised that the low-metallicity nature of the SFRs in the solar neighbourhood may reflect the composition of the giant molecular cloud complex that gave birth to all these associations in a common and wide spread star formation episode. In fact, most of the nearby and young associations are likely to be part of a complex of stars called the Gould Belt \citep{Poppel97,deZeeuw99,Elias09}, thus they may share the same origin (but see \citealt{Bouy15} for a different interpretation of this structure).
However, the present sample includes three star forming regions at larger distances from the Sun and with no relationship with the Gould Belt. Figure~\ref{gradients_obs} indicates that these three regions, in particular the two innermost ones, have [Fe/H] content 0.10-0.15 dex lower than the locus of most of the older clusters at similar Galactocentric distances$\footnote{We note, however, the position of the 500~Myr old NGC4815, the metallicity of which is lower than that of the other intermediate age clusters and more in agreement with the younger clusters. As mentioned by Jacobson et al., this is the youngest among the intermediate-age clusters at that Galactocentric distance. Hence, its lower [Fe/H] may also support an inverse age-metallicity relationship (see also Section~\ref{comp}).}$. This suggests that the lack of metal-rich young clusters may be related to a more general process working on a Galactic scale. 

The second feature regards the evolution of the radial metallicity gradient over time. Whilst the old cluster metallicity distribution clearly shows the presence of a negative gradient of $-$0.10$\pm$0.02 dex~kpc$^{-1}$ between $\sim$5.5 and 8~kpc, as discussed by \citet{Jacobson16}, the distribution of the younger objects suggests a much shallower gradient: we derive slopes equal to $-0.010\pm 0.006$ and $-0.011\pm 0.012$ dex~kpc$^{-1}$ for the SFRs and all the associations younger than 100~Myr (i.e. SFRs + YOCs), respectively, through a linear fit based on the orthogonal distance regression method and taking into account the error bars on both the axes. 
In other words, comparison with the old clusters analysed by Jacobson et al. suggests a flattening of the gradient at very young ages. 

The results on the evolution of the Galactic metallicity distribution is still based on a small number of metallicity determinations and limited to clusters covering a range of $\sim$2~kpc in R$_{gal}$ around the Solar location, thus they cannot be taken as conclusive. However, it is in agreement with and, crucially, extends to younger ages, previous estimates of the evolution of the radial metallicity gradient based on the study of open clusters \citep{Carraro98,Friel02,Magrini09,Andreuzzi11,Cunha16}. On the other hand, this result is at variance with other studies based on observations of field stars (e.g. \citealt{Nordstrom04,Casagrande11,Anders16}). As mentioned already, however, the main difficulty of using field stars to trace the chemical pattern of the Galaxy and its evolution over time is that their ages and distances are not as well constrained as for open clusters, and that field star samples are normally older than 1 Gyr, thus not allowing the study of the latest phases of disc evolution. 
Whilst the Gaia mission \citep{Perryman01,Gaia16} and asteroseismology are allowing and will continue to allow a big step forward in the determination of the distances and ages of field stars, the second problem still remains a major and intrinsic limitation.

\subsection{Comparison with literature results and the metallicity distribution at young ages from other tracers}
In a recent paper \citet{Netopil16} investigated the metallicity 
distribution of a sample of 172 clusters, using a homogenised compilation from the literature. They suggested that young clusters (defined in that paper as those younger than 500~Myr) may be characterised by lower  metallicities than the older ones, at least in the region between 7 and 9 kpc from the Galactic centre; at the same time, they
 confirmed the presence of a negative gradient for these clusters. We mention, however, that their sample did not
include any clusters younger than $\sim 100$~Myr located in the inner Galaxy. 

\begin{figure}
\includegraphics[width=9cm]{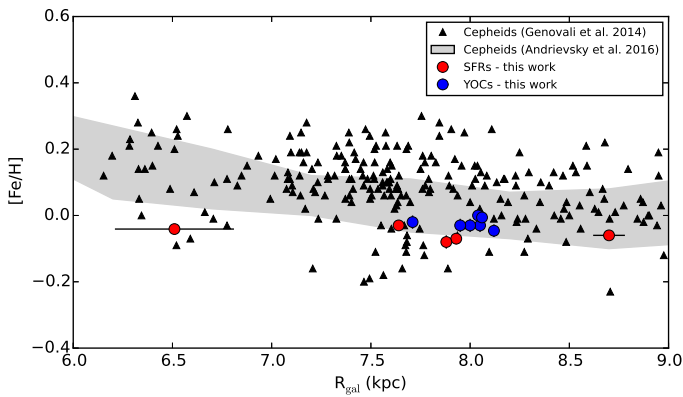} 
\caption{Comparison between the metallicity distributions of SFRs and YOCs by our analysis (red and blue dots, respectively) and those of Cepheids (triangles and grey area). The black triangles represent the values reported in Tables 3 and 4 of \citet{Genovali14}. The grey coloured area contains the bulk of the Cepheids distribution discussed in \citet{Andrievsky16}.}
\label{tracers_cepheids}
\end{figure}

A negative gradient (i.e. $-$0.060$\pm$0.002 dex~kpc$^{-1}$) is also found by \citet{Genovali14} from Cepheids with ages ranging between 20 and 400~Myr and covering a region between 4 and 19~kpc from the Galactic centre. Similar metallicity distributions have been found by other recent investigations based on Cepheids, such as \citet{Andrievsky16,Andrievsky04,Luck11,Pedicelli09,Lemasle08}. The slopes found by these authors (i.e. $-$0.06~$\lesssim$ $\frac{\rm \delta [Fe/H]}{\rm \delta R_{gal}}$ $\lesssim$~$-$0.05    dex~kpc$^{-1}$) are shallower than that found by Jacobson et al. for older clusters (i.e. $-$0.10$\pm$0.02 dex~kpc$^{-1}$), but still not consistent with the values that we obtained for the young associations (i.e. $-0.011\pm 0.012$ dex~kpc$^{-1}$). However, it should be noted that an important parameter 
is the range in Galactocentric distances that they cover. For instance, the Cepheids analysed by Genovali et al. cover a  range of Galactocentric radii (typically from 5 to 15 kpc) that is significantly larger than that of the young associations considered in this work, therefore a strict comparison between the slopes obtained through the two tracers is not entirely appropriate. 
Interestingly, when considering only the Cepheids lying between 6 and 9 kpc from the Galactic centre, as in Fig.~\ref{tracers_cepheids}, we observe that their metallicity distribution is in good agreement with the slope found for the young associations. We also note that the SFRs and YOCs are located close to the lower envelope of the Cepheids distribution. This systematic difference in iron abundance could be the consequence of different choices in the methods of analysis, tools (linelist, atmospheric models), and abundance zeropoints. 
However, this difference could also be a real effect related to the age, since the variable stars are, in average, older than the young associations considered here.


\begin{figure}
\includegraphics[width=9cm]{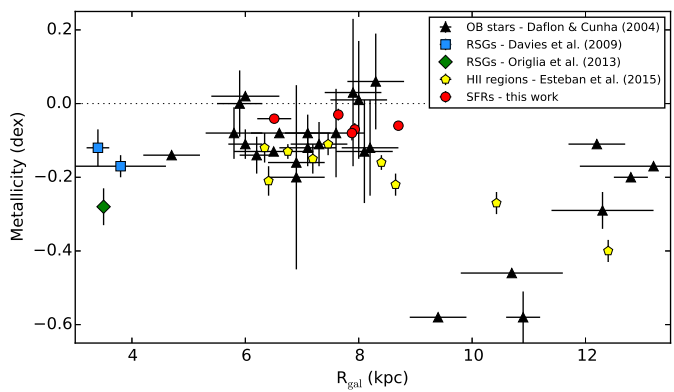} 
\caption{Comparison between the metallicity values obtained for the SFRs by our analysis (in red) and other very young tracers of the present-day radial metallicity gradient: O and B-type stars (black symbols), red supergiants (RSGs; blue and green symbols), and H{\sc ii} regions (yellow symbols). For the RSGs and SFRs we assumed the [Fe/H] values as metallicity determinations, while for the O and B-type stars and H{\sc ii} regions we adopted the log(O)+12 scaled by 8.66~dex, the mean solar abundance of oxygen taken from \citet{Grevesse07}.}
\label{tracers_final}
\end{figure}

Interestingly, the youngest populations in the inner part of the Galaxy,
including O and B-type stars \citep{Daflon04,Nieva11,Nieva12}, red supergiants in clusters and in the Galactic centre
\citep{Davies09a,Davies09b,Martins08,Najarro09,Origlia13,Origlia16}, and H{\sc ii} regions \citep{Rudolph06,Esteban14,Esteban15}, show flatter distributions and close-to-solar metallicities even at small Galactocentric distances. 
Our results confirm these findings based on later type stars in open clusters, the metallicities of which
are in principle easier to constrain than for hot stars or H{\sc ii} regions. 
In Fig. \ref{tracers_final} we plot the distribution of metallicity as a function of Galactocentric distance for the merged sample of different very young tracers, including hot stars, red supergiants, H{\sc ii} regions, and the SFRs in our sample. The datapoints show some scatter, due to the different methods and different abundance scales. 
However, all tracers consistently show that ({\it i}) the inner parts of the Milky Way disc have not undergone a global metal enrichment, but, rather, the innermost young objects appear to share the same metallicity as the others with R$_{\rm gal}$$\gtrsim$6~kpc; ({\it ii}) there is no evidence for any negative metallicity gradient between about 3 and 9 kpc since the distribution in that region is consistent with being flat. This is at variance with HII regions and planetary nebulae observed in nearby spiral galaxies (e.g. \citealt{Magrini16}, and references therein) for which a higher metallicity is observed in the younger populations, but might be similar to what is observed in massive spirals in the CALIFA sample for which, in the inner regions, the metallicity does not increase monotonically but has an inversion assuming a lower value for the youngest stellar populations (S. Sanchez, private communication).


\subsection{Comparison with the models}
\label{comp}
Several chemical and chemo-dynamical evolutionary models have been developed in the past decade making different predictions on the metallicity distribution at different Galactocentric radii, on the slope of the metallicity gradient in the Solar vicinity, and on the evolution of this distribution
with age. Indeed, depending on the assumptions about infall and star formation efficiency, and considering or not a gas density threshold to allow star formation, some of the models predict a flattening of the gradient with
time, while others predict an increase.  A disc formed by pre-enriched gas, and in which a minimum gas density is required to permit the formation of new stars, naturally develops an initial flat metallicity gradient that becomes steeper with time (see e.g. \citealt{Chiappini01}). On the other hand, in models in which the disc is formed by primordial gas and the star formation can proceed at any gas density, the radial metallicity gradient is typically steeper at early times and flattens as the galaxy evolves \citep{Ferrini94,Hou00,Molla05,Magrini09}. Recently, \citet{Gibson13} have examined the role of energy feedback in shaping the time evolution of abundance gradients within a subset of cosmological hydrodynamical disc simulations drawn from the MUGS (McMaster Unbiased Galaxy Simulations; \citealt{Stinson10}) and MaGICC (Making Galaxies in a Cosmological Context; \citealt{Brook12}) samples. The two sets of simulations adopt two feedback schemes: the conventional one in which about 10-40$\%$ of the energy associated with each supernova (SN) is used to heat the ISM, and the enhanced feedback model in which a larger quantity of energy per SN is released, distributed, and recycled over large scales. The resulting time evolution of the radial gradients is different in the two cases: a strong flattening with time in the former and a relatively flat and temporally invariant abundance gradient in the latter. Our results clearly favour those models which do predict the flattening of the gradient with time.

However, according to most models, the global metallicity in the 
inner part of the Galaxy should increase (or, at least not decrease) with time. Whilst an inverse age-metallicity relation (i.e. older stars being more metal rich than young ones) may be expected at the solar radius and beyond, due to radial migration, at inner Galactocentric distances young
stars should not be more metal poor than older ones.
\begin{figure}
\includegraphics[width=9cm]{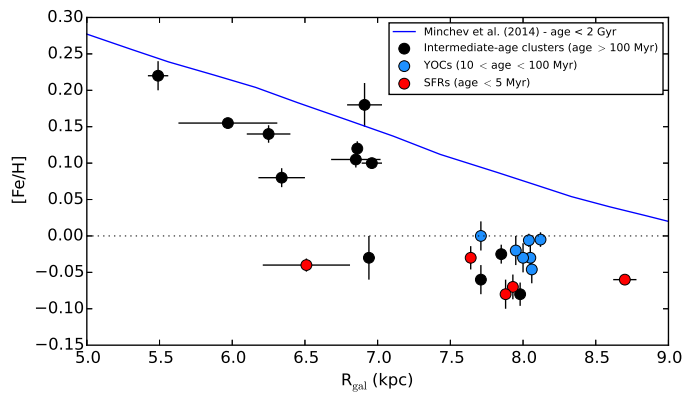} \caption{Radial metallicity distribution of regions younger and older than 100~Myr compared with the model of \citet{Minchev14} for the age interval between 0 and 2 Gyr including radial migration (blue line).}
\label{gradient_minchev}
\end{figure}

\begin{figure}
\includegraphics[width=9cm]{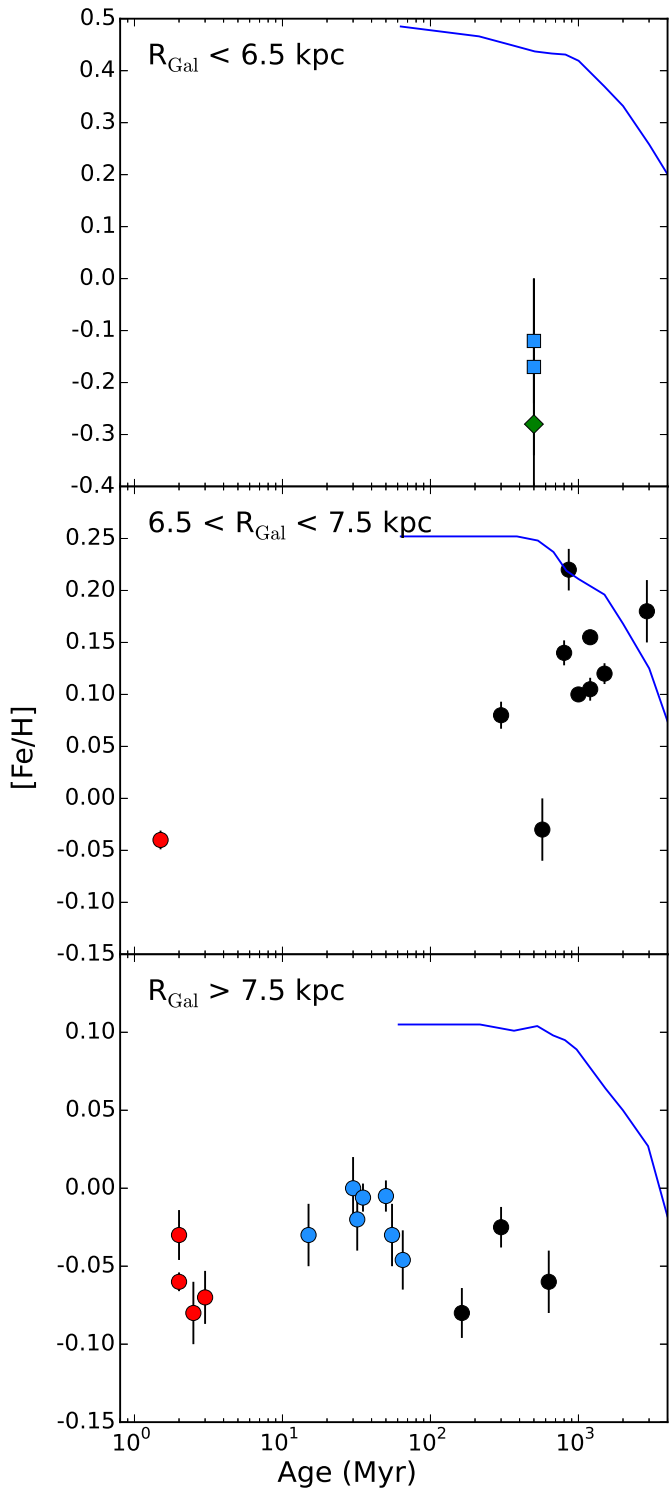} \caption{[Fe/H] variation with age at different Galactic radii. The clusters observed by the Gaia-ESO Survey and the model of \citet{Minchev14} are represented. The colour code is the same as Fig.~\ref{gradient_minchev}.}
\label{FeH_Age}
\end{figure}

As an example, in Fig.~\ref{gradient_minchev}, we show the radial metallicity distribution of the open clusters and young associations, colour-coded by age, and we compare them with the results of the chemical evolution model of \citet{Minchev14} for populations younger than 2~Gyr (no younger bins are available in that paper). The model of \citet{Minchev14} takes into account also the effect of radial migration, which, however, is almost negligible for the youngest population and conversely it is much more important for the oldest and hottest stellar populations. 
As already pointed by \citet{Jacobson16}, both an offset and slope problems are present when comparing our results with the observations: the model predicts slightly higher abundances and a flatter slope than what is observed. For instance, at the solar Galactocentric distance, where several cluster metallicity measurements are available, there are no clusters with [Fe/H]~$=$~0.10~dex as expected by the model, but they all have metallicities ranging from $-$0.1 to 0.0 dex. 

In Fig.~\ref{FeH_Age}, we show the metallicities of the clusters as a function of their age, and again we compare with the results of the model. The data are divided into three panels according to the Galactocentric distance bin: R$_{\rm Gal}$ $<$ 6.5 kpc, 6.5 kpc $<$ R$_{\rm Gal}$ $<$ 7.5 kpc, and R$_{\rm Gal}$ $>$ 7.5 kpc.  We note that the model does not extend to ages below 60 Myr.  However, we might expect a continuous behaviour for the model curves and a constant flat trend from 100 Myr to the present time, starting from the metallicity reached at 100 Myr. 
While for the oldest populations the model and the observations might be reconciled with a rescaling of the slope and with an offset in metallicity, for the youngest population the model curve and the observations are hardly compatible. There is no possibility to explain in the framework of the model the lower metallicity of the youngest population with respect to the oldest one. 
Clearly, radial migration is not the reason why very young clusters in the inner Galaxy have metallicities below those of older ones, as claimed by \citet{Anders16}.


The distribution of metallicity of low-mass stars in very young clusters, along with that of other tracers of the present-day metallicity, instead implies a decrease of the metal content in the inner part of the thin disc in the last few hundreds of Myr; this is likely due to a significantly more complex combination of star formation, accretion history and inflows, radial gas flows, supernova feedback, and other factors. In particular, very recent accretion events may have had a role in shaping the latest evolution of the inner disc. Most obviously these results could be used as additional constraints within existing modellings.

%

\section{Summary and conclusions}
\label{conclusions}
In this paper we employed the analysis products internally released to the Gaia-ESO consortium in the iDR4 catalogue to identify a sample of secure members of ten SFRs and YOCs targeted by the Survey. 
Members of an additional SFR (Chamaeleon I) and a YOC (Gamma Velorum) were previously identified by \citet{Spina14,Spina14b} using previous releases of the Gaia-ESO catalogues. We homogeneously used the [Fe/H] values listed in the iDR4 to determine the median metallicity for each cluster. The main key advantages of the present study are that (i) the median metallicities are based on large samples of G,K-type cluster members; (ii) the [Fe/H] determinations are the result of a homogeneous analysis that allows a meaningful comparison of different populations on the same scale; (iii) for the first time we determined the metal content of three distant SFRs (distance from the Sun $>$ 500 pc), two of which are located in the inner part of the disc.

Our main findings can be summarised as follows:
\begin{itemize}
\item all the YOCs and SFRs analysed in this paper have close-to-solar or slightly sub-solar metallicities. Strikingly, none of them appear to be metal rich. Since our sample of SFRs spans different Galactocentric radii, from $\sim$6.5~kpc to 8.70~kpc, the obvious implication is that the low metal content that characterises these associations is not a peculiarity of the local ISM, as has been previously hypothesised by \citet{Spina14b}; on the contrary, it may be the result of a process of chemical evolution that involved a wide area of the Galactic disc and that influenced the chemical content of the youngest stars regardless of their position within the disc. 
\item The comparison with older clusters (Fig.~\ref{gradients_obs}) suggests that the innermost SFRs at R$_{Gal}$$\sim$7~kpc have [Fe/H] values that are 0.10-0.15~dex lower than the majority of intermediate-age clusters located at similar Galactocentric radii. In addition, while the older clusters clearly trace a negative gradient (i.e. $-$0.10$\pm$0.02 dex~kpc$^{-1}$), the distribution of the youngest objects seems much flatter: $-0.010\pm 0.006$ and $-0.011\pm 0.012$ dex~kpc$^{-1}$ considering SFRs only and all clusters younger than 100~Myr, respectively. Therefore, this comparison between the older and younger stars suggests that the Galactic thin disc experienced further evolution in the last Gyr and a flattening of its metallicity gradient at very young age. However, since these results are based on a small number of young clusters and associations located within 6 and 9 kpc from the Galactic centre, additional metallicity determinations for distant SFRs and YOCs are required to corroborate (or rule out) this scenario.
\item The Cepheids are excellent distance indicators whose ages range between 20 and 400~Myr. The metallicity gradient determined through these variable stars lying between 5 and 15~Gyr from the Galactic centre (e.g. $-$0.060$\pm$0.002 dex~kpc$^{-1}$; \citealt{Genovali14}) is shallower than that obtained through old and intermediate-age clusters, but still steeper than that found from our sample of young associations. However, as shown in Fig.~\ref{tracers_cepheids}, if we consider only the Cepheids located over the same range of Galactocentric distances as the SFRs (i.e. 6$\lesssim$R$_{\rm Gal}$$\lesssim$9~kpc), we observe that their metallicity distribution is in good agreement with the slope found for the young associations.
Many observational studies that used different young populations (e.g. O and B-type stars, red supergiants, HII regions) as metallicity tracers in the inner part of the Galaxy are in agreement with our results (Fig.~\ref{tracers_final}). The flattening of the metallicity gradient is also predicted by some of the models of the chemical evolution of galaxies. On the other hand, other chemo-dynamical models are not able to reproduce the observed feature of the present-day metallicity distribution (see Figs.~\ref{gradient_minchev} and ~\ref{FeH_Age}). To our knowledge, current models do not predict a decrease of the average metallicity in the inner part of the disc, with young clusters being generally more metal-poor than older ones.

\end{itemize}

The chemical composition of the youngest associations in our Galaxy are likely to be the result of a balance of different processes, such as star formation activity and gas flows, that resulted in the chemical evolution of the Milky Way. For this reason, the study of the metal content of these associations and of the present-day metallicity gradient represents an unique approach to achieve an insightful understanding of the elements that govern the history of our Galaxy.

\begin{acknowledgements}
Based on data products from observations made with ESO Telescopes at the La Silla Paranal Observatory under programme ID 188.B-3002. These data products have been processed by the Cambridge Astronomy Survey Unit (CASU) at the Institute of Astronomy, University of Cambridge, and by the FLAMES/UVES reduction team at INAF/Osservatorio Astrofisico di Arcetri. These data have been obtained from the Gaia-ESO Survey Data Archive, prepared and hosted by the Wide Field Astronomy Unit, Institute for Astronomy, University of Edinburgh, which is funded by the UK Science and Technology Facilities Council.
This work was partly supported by the European Union FP7 programme through ERC grant number 320360 and by the Leverhulme Trust through grant RPG-2012-541. We acknowledge the support from INAF and Ministero dell' Istruzione, dell' Universit\`a' e della Ricerca (MIUR) in the form of the grant "Premiale VLT 2012" and PRIN-2014. The results presented here benefit from discussions held during the Gaia-ESO workshops and conferences supported by the ESF (European Science Foundation) through the GREAT Research Network Programme.
L.S. acknowledges the support from FAPESP (2014/15706-9).
\end{acknowledgements}

%
%

\bibliographystyle{aa}
\bibliography{/Users/lspina/Dropbox/papers/bibliography.bib}

\end{document}